\documentclass[aps,pra,preprint,superscriptaddress,showpacs]{revtex4}
\usepackage{amsmath,amssymb,bm,color,graphicx,epsfig}

\newcommand{\ket}[1]{\left| {#1} \right\rangle}
\newcommand{\bra}[1]{\left\langle {#1}\right|}

\begin{document}
\preprint{}
\title{Entanglement, Berry Phases, and Level Crossings for the
Atomic Breit-Rabi Hamiltonian}
\author{Sangchul Oh}
\email{scoh@pks.mpg.de}
\affiliation{Max Planck Institute for the Physics of Complex Systems, 
     Noethnitzer Str. 38, D-01187 Dresden, Germany}
\author{Zhen Huang}
\affiliation{Purdue Univ., Dept. Chem., W Lafayette, IN 47906 USA}
\author{Uri Peskin}
\affiliation{Technion Israel Institute of Technology, Schulich Faculty of 
   Chemistry, IL-32000 Haifa, Israel}
\author{Sabre Kais~\footnote{Corresponding author}}
\email{kais@purdue.edu}
\affiliation{Max Planck Institute for the Physics of Complex Systems, 
     Noethnitzer Str. 38, D-01187 Dresden, Germany}
\affiliation{Purdue Univ., Dept. Chem., W Lafayette, IN 47906 USA}
\date{\today}
\begin{abstract}
The relation between level crossings, entanglement, and Berry phases is 
investigated for the Breit-Rabi Hamiltonian of hydrogen and sodium atoms, 
describing a hyperfine interaction of electron and nuclear spins in a 
magnetic field. It is shown that the entanglement between nuclear and 
electron spins is maximum at avoided crossings. An entangled state encircling
avoided crossings acquires a marginal Berry phase of a subsystem like 
an instantaneous eigenstate moving around real crossings accumulates a
Berry phase. Especially, the nodal points of a marginal Berry phase 
correspond to the avoided crossing points.
\end{abstract}
\pacs{03.65.Ud, 03.65.Vf, 32.60.+i}
\maketitle

\section{Introduction}

Energy levels, eigenvalues of a Hamiltonian, play the most primary role in
determining properties of a quantum system. When energy levels cross
or avoided cross as parameters of a Hamiltonian vary, various interesting
phenomena happen. For example, if two instantaneous energy levels of a 
time-dependent Hamiltonian are avoided crossing, the non-adiabatic tunneling 
called the Landau-Zener tunneling between them takes place~\cite{Landau}. 
Closely related to this, the runtime of adiabatic quantum computation is 
inversely proportional to the square of the energy gap between the ground 
and first exited levels~\cite{Farhi01}. An eigenstate encircling 
adiabatically degeneracy points accumulates a Berry phase in addition to 
a dynamical phase~\cite{Berry84,Shapere}. In quantum chemistry, a conical 
intersection of electronic energy surfaces of molecules plays a key role 
in understanding ultrafast radiationless reactions~\cite{Yarkony96,Kuppermann}.
A quantum phase transition, a dramatic change in a ground state as parameters 
of a system vary, is related with crossings or avoided crossings of
two lowest energy levels~\cite{Vojta03}. Kais {\it et al}. have shown that 
the finite size scaling method can be used for studying the critical behavior, 
i.e., the level degeneracy or absorption, of a few-body quantum Hamiltonian 
$H(\lambda_1,\cdots,\lambda_k)$ as a function of a set of parameters 
$\{\lambda_i\}$~\cite{kais0,kais12}. These parameters could be the external 
fields, inter-atomic distances, nuclear charges for stability of negative 
ions, cluster size, and optical lattice parameters such as the potential 
depth~\cite{kais34}. Thus, it is important to develop a way of finding level 
crossings and to understand how eigenstates or relevant physical quantities 
change at crossing or avoided crossings.

Recently Bhattacharya and Raman presented a powerful algebraic method for 
finding level crossings without solving an eigenvalue problem 
directly~\cite{Bhattacharya06}. Along with this mathematical 
way, it is necessary to understand what physical quantities can be used to 
detect or characterize crossings or avoided crossings. First of all, the 
measurement of a Berry phase could be a good way to detect level crossings 
because it due to level crossings. It is well known that avoided crossings 
or glancing intersections are not the source of Berry phases~\cite{Yarkony96}.
Is there any way that Berry phases can detect {\it avoided level 
crossings} ? Here we show that the marginal Berry phase of an entangled
state could be an indicator to avoided level crossings.

The entropy is an another indicator to level crossings. Since level crossing 
or avoided crossings are accompanied with a drastic change in eigenstates, 
any contents of information on relevant eigenstates may also vary. The Shannon 
entropy of the electron density measures the delocalization or the lack of 
structure in the respective distribution. Thus the Shannon entropy is maximal 
for uniform distribution, that is, for an unbound system, and is minimal when
the uncertainty about the structure of the distribution is minimal~\cite{kais5}.
Gonz\'alez-F\'erez and Dehesa showed that the Shannon entropy could be 
used as an indicator of avoided crossings~\cite{Gonzalez03}. The von Neumann 
entropy, the quantum version of the Shannon entropy, is a good entanglement 
measure for a bipartite pure state.  In quantum information, much attention 
has been paid to the relation between entanglement and quantum phase 
transitions~\cite{Amico08,kais6}. Recently one of the authors investigated  
the relation between entanglement, Berry phases, and level crossings for two 
qubits with the XY-type interaction and found that the level crossing is not 
alway accompanied with the abrupt change in entanglement~\cite{scoh08}. 

In this paper, in order to study how entanglement and Berry phases vary at 
level crossings, we consider the Breit-Rabi Hamiltonian describing a hyperfine 
interaction of electron and nuclear spins in a uniform magnetic 
field~\cite{Breit-Rabi}. It is shown that the von Neumann entropy of the
electron (or nuclear) spin is maximum at avoided crossings. 
It is demonstrated that the significant changes in Berry phases and 
entanglement are closely related to level crossings. 
We show that the marginal Berry phase of the electron (or nuclear) spin 
could be a good indicator to avoided level crossings. 
The marginal Berry phase has nodal points at the avoided crossing points. 

The paper is organized as follows. In Sec.~\ref{section_BR}, the Breit-Rabi
Hamiltonian is introduced. In Sec.~\ref{section_hydrogen}, as a specific
application of the Breit-Rabi Hamiltonian, we consider a hyperfine interaction
between a nuclear spin $1/2$ and an electron spin $1/2$ of a hydrogen atom 
in a magnetic field. We analyze the close relation between entanglement,
Berry phases, and level crossings. In Sec.~\ref{section_sodium}, we make 
an similar analysis for a sodium atom with a nuclear spin $3/2$ and an 
electron spin $1/2$. In Sec.~\ref{section_conclusion}, we summarize the main 
results.

\section{Breit-Rabi Hamiltonian}
\label{section_BR}

Let us consider an atom with a single valence electron in the ground state 
with orbital angular momentum $L=0$. In the presence of a uniform magnetic 
field $B$ in the $z$ direction, its atomic spectrum is described by 
the Breit-Rabi Hamiltonian~\cite{Breit-Rabi}, which is given by the sum of 
the hyperfine interaction between a nuclear spin $\bf I$ and an electron 
spin $\bf S$ and their Zeeman couplings to the magnetic field 
\begin{equation}
H = A\, {\bf I}{\bm \cdot}{\bf S} + (a S_z + b I_z)B \,,
\label{Hamil_BR}
\end{equation}
where $A$ is the hyperfine coupling constant, $a=\gamma_e\hbar$, and 
$b=\gamma_n\hbar$. Here $\gamma_e$ and $\gamma_n$ are the electron and 
nuclear gyromagnetic ratios, respectively. The electron spin operator 
${\bf S}$ and the nuclear spin operator ${\bf I}$ are measured in the unit 
of $\hbar$. 

The Breit-Rabi Hamiltonian~(\ref{Hamil_BR}) is well studied to describe double
resonance in nuclear magnetic resonance~\cite{Slichter} and the muon spin 
rotation in semiconductors~\cite{Patteson88}. Although simple and well
understood, it still continues to provide new insights. Recently Bhattacharya 
and Raman found a new class of invariants of the Breit-Rabi
Hamiltonian~\cite{Bhattacharya06}. As will be shown here, it is a prime example
for showing the close relation between level crossings, entanglement, and 
geometric phases. Also it is related to a Hamiltonian of 
electron spin qubits in quantum dots~\cite{Loss98} where the Heisenberg 
interaction between two electron spins can be turned on and off to 
implement the controlled-not gate.

Before applying the Hamiltonian~(\ref{Hamil_BR}) to specific systems, let us
look at its general properties. If $B =0$, then $H$ commutes with 
both the square of the total spin operator ${\bf J}^2$ and $J_z$, where the 
total spin operator is defined by ${\bf J} \equiv {\bf I} + {\bf S}$. However,
for $B\ne 0$, due to the fact that $a \ne b$, the Hamiltonian~(\ref{Hamil_BR})
no longer commutes with ${\bf J}^2$, but still commutes with $J_z$. So the 
eigenvalue $m$ of $J_z$ is a good quantum number for the Breit-Rabi 
Hamiltonian. With ladder operators, $S_\pm = S_x \pm i S_y$ and 
$I_\pm = I_x \pm i I_y$, the Hamiltonian~(\ref{Hamil_BR}) can be rewritten as
\begin{equation}
H = A\,I_zS_z + \frac{A}{2}(S_{+}I_{-} + S_{-}I_{+}) + B (a S_z + b I_z)\,.
\label{Hamil_BR2}
\end{equation}
Let us use a simple notation $\ket{m_S,m_I}$ to represent the product state 
$\ket{S,m_S}\otimes\ket{I,m_I}$, where $\ket{S,m_S}$ is an eigenstate of
${\bf S}^2$ and $S_z$, and $\ket{I,m_I}$ is an eigenstate of ${\bf I}^2$ and 
$I_z$. The first and third terms in Eq.~(\ref{Hamil_BR2}) give the diagonal 
matrix elements 
\begin{subequations}
\begin{equation}
\bra{m_S\,m_I}H\ket{m_S\,m_I} 
= f(m_S, m_I) = A\,m_S\,m_I  + m_S\,aB + m_I\,bB \,.
\end{equation}
The second term in Eq.~(\ref{Hamil_BR2}) corresponds to the off-diagonal 
matrix elements
\begin{align}
&\bra{m_S'\,m_I'}S_{+}I_{_-}\ket{m_S\,m_I} \nonumber\\ 
&=\sqrt{(S-m_S)(S+m_s+1)}\,\sqrt{(I+m_I)(I-m_I+1)}\,
  \delta_{m_S',m_S+1}\delta_{m_I',m_I-1}\,.
\end{align}
\label{matrix_element}
\end{subequations}
Since $m_S'-m_S =1$ and $m_I' - m_I= -1$ (or vice versa), one has the 
selection rule, $\Delta m = (m_S'+m_I') -(m_S+m_I) =0$, that is, the magnetic
quantum number $m=m_S + m_I$ is conserved. This implies that the 
Hamiltonian~(\ref{Hamil_BR}) is block diagonal in the basis set 
$\{\ket{m_S,m_I}\}$ ordered by $m$. 

\section{The Hydrogen Atom in a Uniform Magnetic Field}
\label{section_hydrogen}

\subsection{Eigenvalues and Eigenstates}

As a simple but real system described by the Hamiltonian~(\ref{Hamil_BR}), 
let us consider the interaction between the nuclear spin $I=1/2$ and the 
electron spin $S=1/2$ of a hydrogen atom in a uniform magnetic field.
Since $H$ commutes with the $z$-component of the total spin operator, 
$J_z = S_z + I_z$, it is convenient to arrange the product basis 
$\{\ket{m_S,m_I}\}$ in the decreasing order of the magnetic quantum number 
$m$ of $J_z$ as 
$\left\{\ket{\tfrac{1}{2},\tfrac{1}{2}}, \ket{\tfrac{1}{2},-\tfrac{1}{2}},
\ket{-\tfrac{1}{2},\tfrac{1}{2}}, \ket{-\tfrac{1}{2},-\tfrac{1}{2}}\right\}$. 
By means of Eqs.~(\ref{matrix_element}), the Breit-Rabi Hamiltonian for a
hydrogen atom can be written in the ordered basis
\begin{align}
H
= \frac{1}{4}
\begin{pmatrix}
A + 2(a+b)B & 0     & 0           & 0 \\
0      & -A+2(a-b)B & 2A          & 0 \\
0      & 2A         & -A -2(a-b)B & 0 \\
0      & 0          & 0           & A -2(a+b)B
\end{pmatrix}\,.
\label{Hamil_Hydrogen}
\end{align}
The Hamiltonian~(\ref{Hamil_Hydrogen}) is block diagonal, so it is 
straightforward to obtain its eigenvalues and eigenvectors.
The subspace of $m=\pm 1$ is spanned by 
$\left\{ \ket{\tfrac{1}{2},\tfrac{1}{2}}, 
         \ket{-\tfrac{1}{2},-\tfrac{1}{2}}\right\}$. 
The block Hamiltonian on this subspace is already diagonal and has its 
eigenvalues and eigenvectors
\begin{subequations}
\label{H_eigen1}
\begin{align}
E_{\pm 1}       &= \frac{A}{4} \pm \frac{1}{2}(a + b)B \,,\\
\ket{E_{\pm 1}} &= \ket{\pm\tfrac{1}{2},\pm\tfrac{1}{2}},
\label{H_eva}
\end{align}
\end{subequations}
where the subscripts `$\pm 1$' in $E_{\pm 1}$ denote the magnetic quantum 
number $m=\pm 1$. The block Hamiltonian with $m=0$ is defined on the subspace 
of $\left\{ \ket{\tfrac{1}{2}, -\tfrac{1}{2}}, 
            \ket{\tfrac{1}{2}, -\tfrac{1}{2}} \right\}$ and is written as 
\begin{align}
H_{m=0} 
&= \frac{1}{4}
   \begin{pmatrix}
   -A +2(a-b)B & 2A \\
    2A         & -A - 2(a-b)B
   \end{pmatrix}\,.
\end{align}
One can interpret the Hamiltonian $H_{m =0}$ as that of a spin in an 
effective magnetic field in the $x$-$z$ plane, ${\bf B}_{\rm eff} 
\equiv (A/2, 0, (a-b)B/2)$. 
The eigenvalues and eigenvectors of $H_{m=0}$ can be written easily as
\begin{subequations}
\label{H_eigen0}
\begin{align}
E_{0}^{\pm} 
&= -\frac{A}{4} \pm \frac{1}{2}\sqrt{(a-b)^2 B^2 + A^2}\,, \\
\ket{E_{0}^{+}} 
&= \phantom{+} \cos\frac{\alpha}{2}\, \ket{ \tfrac{1}{2},-\tfrac{1}{2}} 
            +  \sin\frac{\alpha}{2}\, \ket{-\tfrac{1}{2}, \tfrac{1}{2}}\,,
\label{H_evb} \\
\ket{E_{0}^{-}} 
&= - \sin\frac{\alpha}{2}\, \ket{ \tfrac{1}{2},-\tfrac{1}{2}} 
   + \cos\frac{\alpha}{2}\, \ket{-\tfrac{1}{2}, \tfrac{1}{2}}\,,
\label{H_evc}
\end{align}
\end{subequations}
where $\tan\alpha\equiv \frac{A}{(a-b)B}$. In a weak magnetic field limit 
(so called the Zeeman region), the Zeeman energy is smaller
than the hyperfine coupling. At $B= 0$, i.e., $\alpha = \pi/2$, 
the ground eigenstate $\ket{E_{0}^{-}}$ becomes the singlet state,
$\ket{E_{0}^{-}} = \frac{1}{\sqrt{2}}\left(\ket{-\tfrac{1}{2},\tfrac{1}{2}} -
\ket{\tfrac{1}{2}, -\tfrac{1}{2}} \right)$. In a strong magnetic field called
the Paschen-Back region, the Zeeman couplings are dominant. That is, in   
limit of $B\to \infty$, one has $\alpha \to 0$ and $\ket{E_{0}^{-}} \to
\ket{-\tfrac{1}{2},\tfrac{1}{2}}$.

The eigenvalues and eigenstates of the Breit-Rabi Hamiltonian for a hydrogen 
atom, Eqs.~(\ref{H_eigen1}) and (\ref{H_eigen0}) depend on two parameters: 
the hyperfine constant $A$ and the magnetic field $B$. The hyperfine constant 
$A$ of the hydrogen atom in vacuum is positive. However, if a hydrogen atom
is in an inert gas, the hyperfine constant $A$ could be 
negative~\cite{Foner60}, resembling the spin-spin coupling constant in 
a Heisenberg model. We assume that $A$ as well as $B$ varies and can be 
negative. To this end, $A$ in Eqs.~(\ref{H_eigen1}) and (\ref{H_eigen0}) is 
replaced by $fA$ with $-1\le f\le 1$, so $A$ is still kept the positive 
constant in vacuum. If $f$ is negative, so the hyperfine constant.

\begin{figure}[htpb]
\includegraphics[scale=1.0]{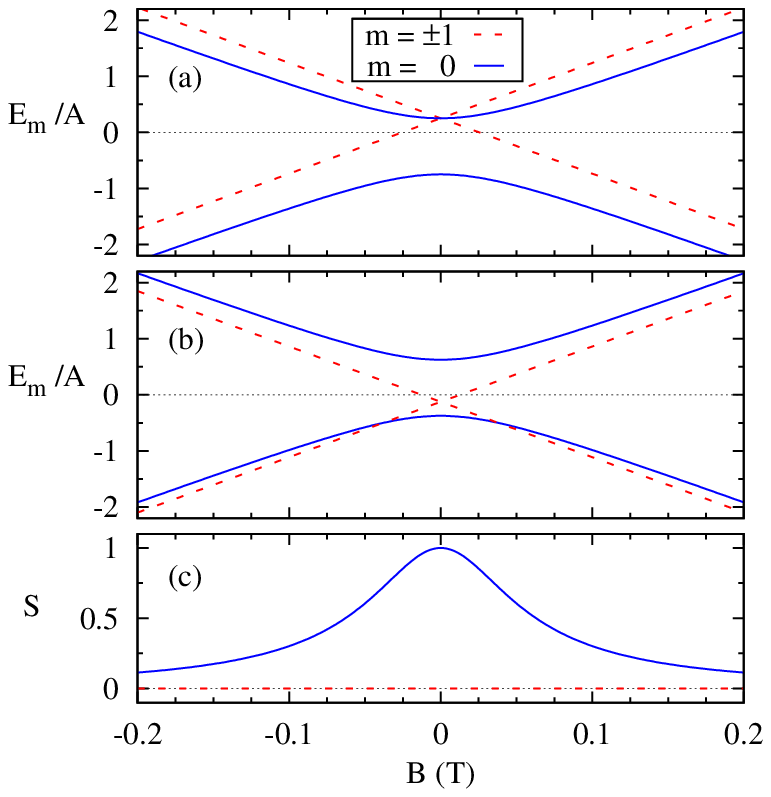}
\caption{(Color online) Energy levels $E_m/A$ of the Breit-Rabi Hamiltonian
for a hydrogen atom as functions of the magnetic field $B$ with (a) $f=1$ and 
(b) $f=-0.5$. (c) the von Neumann entropy $S$ of the electron (or nuclear) spin 
for each eigenstate. Here $a'= 19.767\,\,\text{T}^{-1}$ and 
$b'= -0.03\,\,\text{T}^{-1}$ are taken from Ref.~\cite{Arimondo77}.}  
\label{Fig1}
\end{figure}

\begin{figure}[htbp]
\includegraphics[scale=1.0]{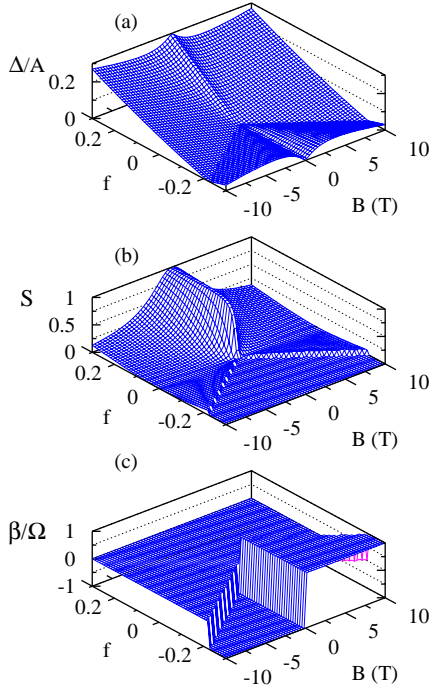}
\caption{(Color online) (a) Energy gap $\Delta/A$ between the ground and first
exited states, (b) the von Neumann entropy $S$ of the electron (or nuclear) 
spin, and (c) the Berry phase $\beta/\Omega$ of the ground state for a hydrogen 
atom as a function of  $f$ and $B$.  Here $a' = 0.01\,\,\text{T}^{-1}$ and 
$b'=-0.1 \,\,\text{T}^{-1}$ are taken to see the jumps clearly at the level 
crossings.} 
\label{Fig2}
\end{figure}

Depending on $f$ and $B$, the ground state of the 
Hamiltonian~(\ref{Hamil_Hydrogen}) is given either by $\ket{E_{\pm 1}}$ or by 
$\ket{E_{0}^{-}}$. It is convenient to plot the energy levels normalized by 
$A$.  Then, Eqs.~(\ref{H_eigen1}) and (\ref{H_eigen0}) become
$E_{\pm1}/A = \frac{f}{4} \pm \frac{1}{2}(a' + b') B$ and 
$E_{0}^{\pm}/A= -\frac{f}{4} \pm \frac{1}{2}\sqrt{(a'-b')^2 B^2 + f^2}$,
where $a' \equiv a/A \approx 19.767\,\,\text{T}^{-1}$ and 
      $b'\equiv b/A \approx -0.03\,\,\text{T}^{-1}$ are taken 
form Ref.~\cite{Arimondo77}, and $B$ is measured in the unit of tesla. 
The energy levels $E_m/A$ are plotted as functions of $B$ for $f=1$ in 
Fig.~\ref{Fig1} (a) and for $f=-0.5$ in Fig.~\ref{Fig1} (b). For $f\ge 0$, 
the ground level is $E_{0}^{-}$. For $f<0$, two levels, $E_{0}^{\pm}$ and
$E_{\pm1}$, with different magnetic quantum numbers cross at  
$f =\frac{2a'b'}{a'-b'} |B|$. Fig.~\ref{Fig2} (a) shows the energy gap 
$\Delta/A$ between the ground and first exited states as a function of $f$ 
and $B$, where we take $a' =0.1\,\,\text{T}^{-1}$ and 
$b' = -0.01\,\,\text{ T}^{-1}$ to see clearly the phase diagram of 
the ground state of the Hamiltonian~(\ref{Hamil_Hydrogen}) determined by 
the magnetic quantum number $m$. As shown in Fig.~\ref{Fig2} (a), the energy
gap $\Delta/A$ vanishes along the lines defined by 
$f = \frac{2a'b'}{a'-b'} |B|$ 
and the negative $f$ axis.  In the region of $f <\frac{2a'b'}{a'-b'} |B|$, 
the ground state becomes either $\ket{E_{+1}}$ or $\ket{E_{-1}}$ with the 
magnetic quantum number $m=1$ or $m=-1$, respectively. On the other hand, 
the ground state in the region defined by $f >\frac{2a'b'}{a'-b'} |B|$ is 
given by $\ket{E_{0}^{-}}$ with the magnetic quantum number $m=0$.
One can see that the magnetic quantum number $m$ of the ground state 
changes abruptly at the level crossing points.

\subsection{Entanglement}

Let us discuss the relation between level crossings and entanglement. 
Entanglement refers to the quantum correlation between subsystems and has no 
classical analog~\cite{book-Gruska,kais7}. When level crossing happens as the 
parameter of the Hamiltonian varies, the ground state changes drastically. 
Entanglement as a physical quantity may also undergo a significant change. 
However, entanglement is not always a good indicator to level crossing as 
shown in Ref.~\cite{scoh08}.

First, let us examine the relation between entanglement and level crossings 
for each eigenstate. The von Neumann entropy $S$ of a subsystem is a good 
entanglement measure for a pure bipartite system. If $\ket{\psi_{AB}}$ is 
a quantum state of a system composed of two subsystems $A$ and $B$, the 
entanglement between $A$ and $B$ is measured by the von Neumann entropy of 
the subsystem,
$S(\rho_A) = -{\rm tr}(\rho_A\log\rho_A)= S(\rho_B) 
           = -{\rm tr}(\rho_B\log\rho_B)$, where the reduced density matrix 
$\rho_A$ of the subsystem $A$ is obtained by tracing out the degrees of 
freedom of $B$ as $\rho_A ={\rm tr}_B( \ket{\psi_{AB}} \bra{\psi_{AB}} )$.     
If the ground state is given by $\ket{E_{\pm 1}}$, i.e., a product state, 
then the von Neumann entropy $S$ of the electron (or nuclear) spin 
is zero. On the other hand, for the quantum state $\ket{E_0^{\pm}}$ of 
the electron and nuclear spins, the von Neumann entropy of the electron 
(or nuclear) spin can be written as
\begin{align}
S(\rho_A) = -\frac{1+\cos\alpha}{2}\,\log_2\frac{1+\cos\alpha}{2}
            -\frac{1-\cos\alpha}{2}\,\log_2\frac{1-\cos\alpha}{2}\,.
\end{align}
Fig.~\ref{Fig1} (c) shows the von Neumann entropy of the electron (or nuclear) 
spin for each eigenstate as a function of $B$. For the eigenstates
$\ket{E_{0}^{\pm}}$, it is maximum at $B=0$, i.e., at the avoided crossing
point. This is analogous to the sharp change in Shannon entropy at avoided 
crossing in Ref.~\cite{Gonzalez03}.

Now, let us look at how entanglement changes at level crossings as the
parameters of the Hamiltonian vary. Fig.~\ref{Fig2} (b) plots the von Neumann 
entropy $S$ of the electron (or nuclear ) spin for the ground state as 
a function of $f$ and $B$. Across the level crossing line, 
$f=\frac{2a'b'}{a'-b'}|B|$, the von Neumann entropy changes abruptly. 
For $f>0$, $S$ becomes 1 as $B$ goes to 0. Along the line of $f=0$, 
the von Neumann entropy $S$ vanishes even though there is no level crossing. 

\subsection{Berry Phase}
\label{hydrogen_berry}

An instantaneous eigenstate encircling the energy level crossing points 
acquires the Berry phase in addition to the dynamical phase. The information
on the level crossings is encoded in the Berry phase. At $B=0$ and $f=1$, the 
two levels $E_{\pm1}$ cross and the other two levels $E_{0}^{\pm}$ are avoided 
crossing. Also $E_{\pm1}$ and $E_{0}^{-}$ cross at $f=\frac{2a'b'}{a'-b'}|B|$. 
Here we focus on the Berry phase due to the level crossing or avoided crossing 
at $B=0$.

Due to the fact that $a\gg |b|$, an electron spin rotates much faster than a 
nuclear spin. We assume the magnetic field ${\bf B}$ is rotated slowly enough
for both the electron and nuclear spins to evolve adiabatically. The magnetic
field ${\bf B} = B\,{\bf\hat{n}}$ in the direction of ${\bf\hat{n}} = 
(\sin\theta\cos\phi, \sin\theta\sin\phi,\cos\theta)$ is constructed starting 
from ${\bf B} = B\,{\bf \hat{z}}$.  First, it is rotated about the $y$ axis by 
angle $\theta$. And it is subsequently rotated about the $z$ axis by angle 
$\phi$. By applying SU(2) rotations corresponding to the above SO(3) rotations 
on the Hamiltonian~(\ref{Hamil_BR}), one obtains the Breit-Rabi Hamiltonian in 
the magnetic field ${\bf B} = B\, {\bf\hat{n}}$
\begin{align}
H(\theta,\phi) 
= A\,{\bf I}{\bm \cdot}{\bf S} 
+ a\,{\bf B}{\bm \cdot}{\bf S} + b\,{\bf B}{\bm \cdot}{\bf I}\,.
\label{Hamil_Berry}
\end{align}
The hyperfine interaction $A\,{\bf I}{\bm \cdot}{\bf S}$ is spherical
symmetric, so the eigenvalues and eigenvectors of the 
Hamiltonian~(\ref{Hamil_Berry}) are identical to those of the 
Hamiltonian~(\ref{Hamil_Hydrogen}) except replacing 
$\ket{\pm\frac{1}{2}}$ by $\ket{{\bf \hat{n}};\pm\frac{1}{2}}$.
Here $\ket{{\bf \hat{n}};\pm\frac{1}{2}}$ are eigenstates of 
${\bf\hat{n}}{\bm \cdot}{\bf S}$ or ${\bf\hat{n}}{\bm \cdot}{\bf I}$. 
If the magnetic field ${\bf B}$ is rotated slowly about the $z$ axis by $2\pi$ 
to make a cone with a solid angle $\Omega = 2\pi(1-\cos\theta)$, then the
instantaneous eigenstate $\ket{{\bf \hat{n}};\pm\frac{1}{2}}$ follows it and
accumulates the Berry phase $\beta_\pm =\mp\frac{1}{2}\Omega$. The total Berry 
phase $\beta$ of electron and nuclear spins is the sum of two phases acquired 
by each one. It depends on the magnetic quantum number $m$ 
\begin{align}
\beta = \left\{ \begin{array}{cl}
                \mp\Omega \quad &\text{for $m=\pm 1$} \,,\\
                0         \quad &\text{for $m=0$} \,.
                \end{array} \right.
\end{align}
As expected, the Berry phase is nonzero only for real crossings, i.e., 
$m=\pm1$. Fig.~\ref{Fig2} (c) plots the total Berry phase as a function of 
$B$ and $f$ and shows that the total Berry phase jumps at the level crossings.
The zero Berry phase of the eigenstates $\ket{E_{0}^{\pm}(\theta,\phi)}$ can 
be understood in two ways. First, two levels $E_{0}^{\pm}$ are avoided 
crossing at $B =0$, so it is zero. Another view is as follows.
Since $\ket{E_{0}^{\pm}}$ is a superposition of 
$\ket{\tfrac{1}{2},-\tfrac{1}{2}}$ and $\ket{-\tfrac{1}{2},\tfrac{1}{2}}$, 
the Berry phase of the electron spin is opposite to that of the nuclear spin 
and they cancel each other. 
	  
Although the entangled states $\ket{E_{0}^{\pm}}$ of electron and nuclear spins 
accumulates no Berry phase, each subsystem (electron spin or nuclear spin) can 
get nonzero marginal Berry phases of mixed states. Following the studies on
geometric phase of mixed states~\cite{Sjoqvist00,Sjoqvist05} and the relation 
between entanglement and marginal Berry phases~\cite{Yi04a,Yi04b,Sjoqvist05}, 
we investigate the relation between avoided level crossings, marginal Berry 
phases, and entanglement. For an adiabatic cyclic evolution parameterized by
$\bf x$, an instantaneous eigenstate of a bipartite system $AB$ can be 
expressed in a Schmidt decomposition $\ket{\psi({\bf x})} = \sum_{i=1}^{M}
\sqrt{p_i} \ket{e_i({\bf x})} \otimes\ket{f_i({\bf x})}$, where 
$\{\ket{e_i({\bf x})}\}_{i=1}^{N_A}$ is an orthonormal basis for a subsystem
$A$, $\{\ket{f_i({\bf x})}\}_{i=1}^{N_B}$ for a subsystem $B$, 
$M\le\min\{N_A,N_B\}$, and $\sum_{i=1}^M p_i =1$. Here our attention is
restricted to the case that the Schmidt coefficients $\sqrt{p_i}$ are
independent of ${\bf x}$. After an adiabatic cyclic evolution implemented by
${\bf x}(0) = {\bf x}(T)$, the total Berry phase of the bipartite system $AB$ 
is given by 
\begin{align}
\beta=\sum_{i=1}^{M} p_i \left(\beta^{A}_{i} + \beta^{B}_{i}\right)\,,
\label{berry_sum}
\end{align}
where $\beta^{A}_{i} = i\oint_C d{\bf x}{\cdot} \bra{e_i({\bf x})} 
                       \nabla_{\bf x} \ket{e_i({\bf x})}$. Then the marginal 
mixed state Berry phase $\Gamma_A$ of a subsystem $A$ is defined by 
\begin{align}
\Gamma_A =\arg\sum_{i}p_i \exp\left(i\beta^A_i\right)\,.
\label{berry_mixed}
\end{align}

\begin{figure}[htbp]
\includegraphics[scale=1.0]{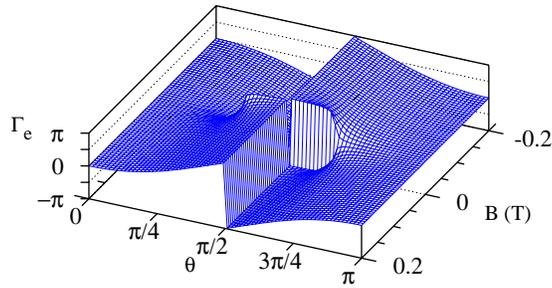}
\caption{(Color online) Marginal Berry phase $\Gamma_e$ of the electron spin 
for the entangled eigenstate $\ket{E_{0}^{-}}$ of a hydrogen atom as a function 
of $B$ and the azimuthal angle $\theta$. Here we take $f=1$, $a' =
19.767\,\,\text{T}^{-1}$, and $b'=-0.03 \,\,\text{T}^{-1}$.}
\label{Fig3}
\end{figure}

With Eqs.~(\ref{berry_sum}) and~(\ref{berry_mixed}), let us analyze how the 
total Berry phase and the marginal Berry phase of $\ket{E_0^{-}}$ depend on
$B$. The two Schmidt coefficients are given by $p_1 = \sin^2\tfrac{\alpha}{2}$ 
and $p_2=\cos^2\tfrac{\alpha}{2}$. It is easy to obtain the marginal 
Berry phase of the electron spin 
$\Gamma_e = \arctan\left(\cos\alpha\tan\frac{\Omega}{2}\right)$
and the average Berry phase of the electron spin 
$\beta_e \equiv p_1\beta_1^e + p_2 \beta_2^e =\frac{\Omega}{2}\cos\alpha$. 
In the limit of $B\gg1$, i.e., $\alpha\to 0$, one has $\ket{E_0^{-}} \to
\ket{-\tfrac{1}{2},\tfrac{1}{2}}$ and $\beta_e={\Omega}/{2}$. Also the marginal
Berry phase of the electron spin is given by $\Gamma_e = {\Omega}/{2}$ for 
$0\le \theta <\frac{\pi}{2}$.  Fig.~\ref{Fig3} plots $\Gamma_e$
as a function of $B$ and the azimuthal angle $\theta$. The marginal Berry 
phase of the electron spin jumps at $\theta=\pi/2$ and $B=0$.
The node at $B=0$ corresponds to the avoided crossing. 

\section{The Sodium  Atom in a Uniform Magnetic Field}
\label{section_sodium}

\subsection{Energy Spectrum}
Now we consider an $^{23}$Na atom in its $3S_{1/2}$ ground state in the 
presence of a uniform magnetic field $B$ along the $z$ axis. The nuclear 
and electron spins of an $^{23}$Na atom are $I=2/3$ and $S=1/2$, respectively.
As in Sec.~\ref{section_hydrogen}, it is convenient to arrange 
the product basis $\{\ket{m_S,m_I}\}$ in the decreasing order of 
the magnetic quantum number $m$ of $J_z$ as follows.
$\left\{\ket{\frac{1}{2},\frac{3}{2}} \right\}$,
$\left\{\ket{\frac{1}{2},\frac{1}{2}}, \ket{-\frac{1}{2},\frac{3}{2}}\right\}$,
$\left\{\ket{\frac{1}{2},-\frac{1}{2}},\ket{-\frac{1}{2},\frac{1}{2}}\right\}$, 
$\left\{\ket{\frac{1}{2},-\frac{3}{2}},\ket{-\frac{1}{2},-\frac{1}{2}}\right\}$,
and $\left\{\ket{-\frac{1}{2},-\frac{3}{2}} \right\}$. For example, 
$\left\{ \ket{\frac{1}{2},\frac{1}{2}}, \ket{-\frac{1}{2},\frac{3}{2}}\right\}$ 
spans the subspace of $m=1$. In this ordered basis set, the 
Hamiltonian~(\ref{Hamil_BR}) for the sodium atom  can be represented by 
a block-diagonal matrix
\begin{align}
H = 
\begin{pmatrix}
f({\frac{1}{2},\frac{3}{2}}) & 0 & 0 & 0 & 0 & 0 & 0 & 0 \\
0 & f(\frac{1}{2},\frac{1}{2}) & \frac{\sqrt{3}}{2}A & 0 & 0 & 0 & 0 & 0 \\
0 &\frac{\sqrt{3}}{2}A & f(\frac{-1}{2},\frac{3}{2}) & 0 & 0 & 0 & 0 & 0 \\
0 & 0 & 0 & f(\frac{1}{2},\frac{-1}{2}) & \frac{A}{2} & 0 & 0 & 0 \\
0 & 0 & 0 & \frac{A}{2} & f(\frac{-1}{2},\frac{1}{2}) & 0 & 0 & 0 \\
0 & 0 & 0 & 0 & 0 & f(\frac{1}{2},\frac{-3}{2}) & \frac{\sqrt{3}}{2}A & 0 \\
0 & 0 & 0 & 0 & 0 & \frac{\sqrt{3}}{2}A & f(\frac{-1}{2},\frac{-1}{2}) & 0 \\
0 & 0 & 0 & 0 & 0 & 0 & 0 & f(\frac{-1}{2},\frac{-3}{2})  
\end{pmatrix},
\label{Hamil_Sodium}
\end{align}
where $f(m_S,m_I) \equiv A\,m_S\,m_I + m_S\,aB + m_I\,bB$.
Each block is at most a $2\times 2$ matrix and  can be easily diagonalized.
First, consider the subspace of $m = \pm 2$. The corresponding eigenvalues 
and eigenvectors can be written as
\begin{subequations}
\label{Na_m2}
\begin{align}
E_{\pm 2} &= \frac{3}{4}A \pm \frac{1}{2}(a+3b)B\,,\\
\ket{E_{\pm 2}} &= \ket{\pm\tfrac{1}{2},\pm\tfrac{3}{2}} \,. 
\label{Na_ev2}
\end{align}
\end{subequations}
Notice that Eqs.~(\ref{Na_m2}) are comparable to Eqs.~(\ref{H_eigen1}).
Second, in the subspace with $m= 1$ spanned by 
$\left\{\ket{\frac{1}{2},-\frac{1}{2}}, \ket{-\frac{1}{2},\frac{1}{2}}\right\}$, 
one obtains the eigenvalues and eigenvectors,
\begin{subequations}
\label{Eq_plus}
\begin{align}
E_{+1}^{\pm} 
&= -\frac{A}{4} + bB \pm \frac{1}{2}\sqrt{\bigl(A + (a-b)B\bigr)^2 + 3A^2}\,,\\
\ket{E_{+1}^+} 
&= \phantom{+} \cos\frac{\alpha_1}{2}\, \ket{ \tfrac{1}{2},\tfrac{1}{2}} 
            +  \sin\frac{\alpha_1}{2}\, \ket{-\tfrac{1}{2},\tfrac{3}{2}}\,,
\label{Na_p1p}\\
\ket{E_{+1}^-}&= -\sin\frac{\alpha_1}{2}\, \ket{ \tfrac{1}{2},\tfrac{1}{2}} 
            +  \cos\frac{\alpha_1}{2}\, \ket{-\tfrac{1}{2},\tfrac{3}{2}}\,,
\label{Na_p1m}
\end{align}
\end{subequations}
where $\tan\alpha_1\equiv \frac{\sqrt{3} A}{A + (a-b)B}$. 
Third, the Hamiltonian of $m=-1$ is defined on the subspace 
 spanned by $\left\{ \ket{\frac{1}{2},-\frac{3}{2}}, 
\ket{-\frac{1}{2},-\frac{1}{2}} \right\}$. 
Its eigenvalues and eigenstates are given by
\begin{subequations}
\label{Eq_minus}
\begin{align}
E_{-1}^{\pm} &= -\frac{A}{4} -bB 
             \pm \frac{1}{2}\sqrt{\bigl(A - (a-b)B\bigr)^2 + 3A^2}\,,\\
\ket{E_{-1}^+} 
&= \phantom{+} \cos\frac{\alpha_2}{2}\, \ket{-\tfrac{1}{2},-\tfrac{1}{2}} 
            +  \sin\frac{\alpha_2}{2}\, \ket{ \tfrac{1}{2},-\tfrac{3}{2}}\,,
\label{Na_m1p} \\
\ket{E_{-1}^-} 
&= - \sin\frac{\alpha_2}{2}\, \ket{-\tfrac{1}{2},-\tfrac{1}{2}} 
   + \cos\frac{\alpha_2}{2}\, \ket{ \tfrac{1}{2},-\tfrac{3}{2}} \,,
\label{Na_m1m}
\end{align}
\end{subequations}
where $\tan\alpha_2\equiv \frac{\sqrt{3}A}{A -(a-b)B}$. Note that
Eqs.~(\ref{Eq_minus}) can be obtained from Eqs.~(\ref{Eq_plus}) by replacing 
$B$ with $-B$. Finally, the subspace of $m=0$ is spanned by 
$\left\{\ket{\frac{1}{2},-\frac{1}{2}},\ket{\frac{1}{2},-\frac{1}{2}}\right\}$. 
The corresponding eigenvalues and eigenvectors are given by
\begin{subequations}
\label{Na_ev0}
\begin{align}
E_{0}^{\pm} &= -\frac{A}{4} \pm \frac{1}{2}\sqrt{(a-b)^2B^2 + 4A^2} \,,\\
\ket{E_0^+} &= 
\phantom{+} \cos\frac{\alpha_0}{2}\, \ket{ \tfrac{1}{2},-\tfrac{1}{2}} 
         +  \sin\frac{\alpha_0}{2}\, \ket{-\tfrac{1}{2}, \tfrac{1}{2}} \,, 
\label{Na_0p} \\
\ket{E_0^-} &= 
        -   \sin\frac{\alpha_0}{2}\, \ket{ \tfrac{1}{2},-\tfrac{1}{2}} 
	+   \cos\frac{\alpha_0}{2}\, \ket{-\tfrac{1}{2}, \tfrac{1}{2}}\,,
\label{Na_0m}
\end{align}
\end{subequations}
where $\tan\alpha_0\equiv \frac{A}{(a-b)B}$. As expected, Eqs.~(\ref{Na_ev0})
is very similar to Eqs.~(\ref{H_eigen0}) in the case of a hydrogen atom.

\subsection{Entanglement}

\begin{figure}[htbp]
\includegraphics[scale=1.0]{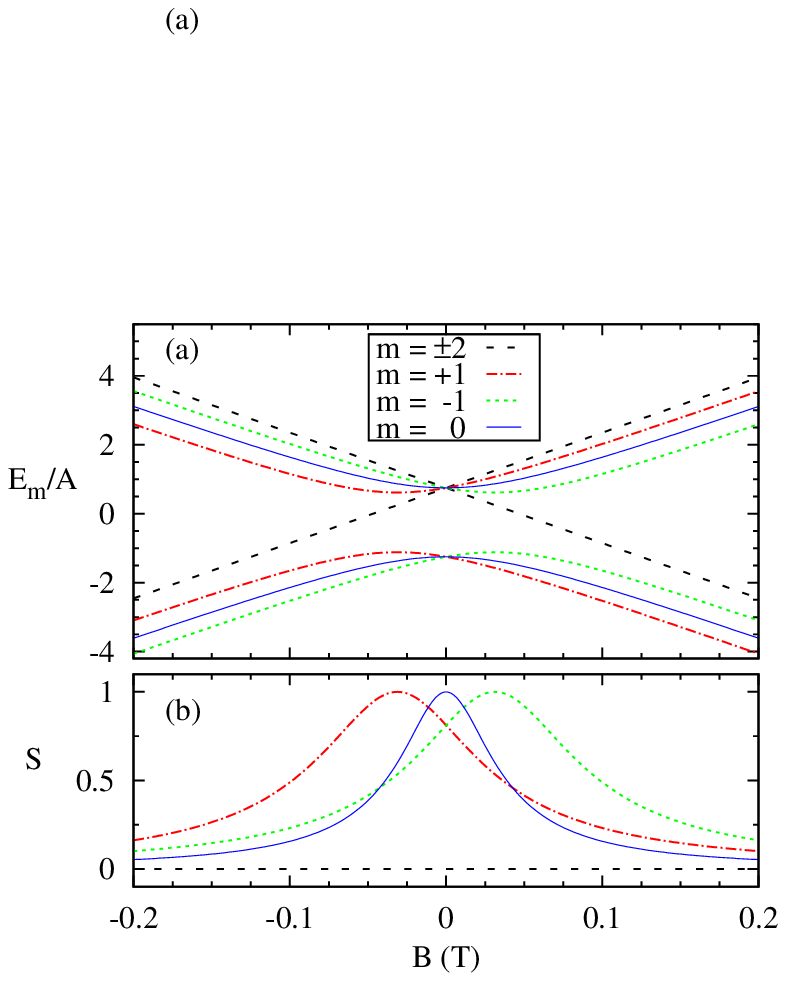}
\caption{(Color online) (a) Energy levels $E_{m}/A$ and (b) the von Neumann
entropy $S$ of the electron (or nuclear) spin for the eigenstates of the sodium
atom as a function of $B$. The two parameters of a sodium atom, 
$a' =32.091\,\,\text{T}^{-1}$ and $b'=-0.012709 \,\,\text{T}^{-1}$ are taken 
from Ref.~\cite{Arimondo77}.} 
\label{Fig4}
\end{figure}

Let us examine the relation between entanglement and level crossings 
or avoided crossings for a sodium atom. 
With the values of the parameters $A$, $a$, and $b$ of the $^{23}$Na 
atom in Ref.~\cite{Arimondo77}, energy levels $E_{m}^{\pm}/A$ are plotted
in Fig.~\ref{Fig4} (a). The von Neumann entropies of the electron 
(or nuclear) spin for each eigenstates are shown in Fig.~\ref{Fig4} (b).
The ground state is given by $\ket{E_{+1}^{-}}$ for $B>0$ and 
$\ket{E_{-1}^{-}}$ for $B<0$. Two levels, $E_{+1}^{+}$ and 
$E_{+1}^{-}$, are avoided crossing and maximally entangled at 
$A -(a-b)B= \sqrt{3}A$.  Another two levels, $E_{-1}^{+}$ and $E_{-1}^{-}$, 
are avoided crossing and maximally entangled at $A + (a-b)B= \sqrt{3}A$.
Two levels with $m=0$, $E_{0}^{\pm}$ are avoided crossing and maximally
entangled at $B=0$. Two levels $E_{\pm2}$ show real crossing at $B=0$ and have
zero von Neumann entropies. Again one can see that the eigenstate is maximally
entangled at the avoided crossing point. This is analogous to the results 
in Ref.~\cite{Gonzalez03}, where Shannon entropy is used as an indicator of
avoided crossings.

\subsection{Berry phase}

As in Sec.~\ref{hydrogen_berry}, let us consider an adiabatic cyclic evolution
of nuclear and electron spins of a sodium atom by rotating the magnetic 
field $B\,{\bf\hat{n}}$ slowly. For a adiabatic rotation keeping the azimuthal 
angle $\theta$ constant and varying the polar angle $\phi$ from $0$ to $2\pi$,
the instantaneous eigenstates accumulates the total Berry phases proportional 
to the magnetic quantum number $m$, $\beta = \mp m \Omega$. In contrast to a
hydrogen atom, the ground state is given either by $\ket{E_{+1}^{-}}$ or by
$\ket{E_{-1}^{-}}$ with $m=\pm 1$, so it acquires the total Berry phase $\beta
=\mp\Omega$.

\begin{figure}[htbp]
\includegraphics[scale=1.0]{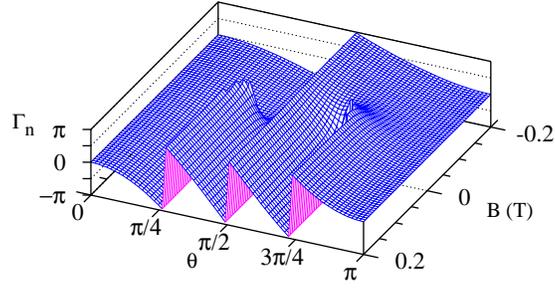}
\caption{(Color online) Marginal Berry phase of the nuclear spin $\Gamma_n$ for
   the eigenstate $\ket{E_{+1}^{-}}$  as 
a function of $B$ and $\theta$.}
\label{Fig5}
\end{figure}

Let us analyze how the marginal Berry phase of the entangled state is related 
to the avoided crossings. We focus on the eigenstate, $\ket{E_{+1}^{-}}$. It has 
two Schmidt coefficients, 
$p_1 = \sin^2\tfrac{\alpha_1}{2}$ and $p_2=\cos^2\tfrac{\alpha_1}{2}$.
With Eq.~(\ref{berry_sum}), one obtains  the total phase as a sum of the Berry 
phases acquired by nuclear and electron spins with weights of the Schmidt 
coefficients,
\begin{align}
\beta = \sin^2\tfrac{\alpha_1}{2}
        \left( -\tfrac{\Omega}{2} - \tfrac{\Omega}{2}\right)
      + \cos^2\tfrac{\alpha_1}{2}
        \left(+\tfrac{\Omega}{2}  - \tfrac{3\Omega}{2}\right)
      = -\Omega\,.
\end{align}
From Eq.~(\ref{berry_mixed}), one obtains the marginal Berry phases of an
electron spin $\Gamma_e$ and of nuclear spin $\Gamma_n$
\begin{subequations}
\begin{align}
\Gamma_n &= \arg\bigl[\,
             \sin^2\tfrac{\alpha_1}{2}\,e^{-i\Omega/2} 
	   + \cos^2\tfrac{\alpha_1}{2}\,e^{-i3\Omega/2}\, \bigr] \,,\\
\Gamma_e &=\arctan\left[\,\cos\alpha_1\tan\tfrac{\Omega}{2}\,\right]\,.
\end{align}
\end{subequations}
Fig.~\ref{Fig5} plots the marginal Berry phase of a nuclear spin $\Gamma_n$ 
as a function of $B$ and the azimuthal angle $\theta$. In the limit of 
$B\gg 1$, i.e., $\alpha_1\to 0$, one has $\ket{E_{+1}^{-}} \to
\ket{-\tfrac{1}{2}, \tfrac{3}{2}}$, $\Gamma_e=\Omega/2$, and
$\Gamma_n=-3\Omega/2$. It is clearly seen that 
the node of the marginal Berry phase of a nuclear (or electron) spin 
corresponds to the avoid crossing at $A + (a-b)B = \sqrt{3}A$. Thus it could 
be expected that the marginal Berry phase of a subsystem for an entangled 
state has a node at avoided crossings.

\section{Conclusions}
\label{section_conclusion}

We have considered the Breit-Rabi Hamiltonians for hydrogen and sodium atoms, 
describing the hyperfine interaction between a nuclear spin and an electron 
spin in the presence of a magnetic field. We have examined the relation between
level crossings, entanglement, and Berry phases. It is shown that entanglement 
between nuclear and electron spins is maximum at avoided crossing points. 
The Berry phase and the von Neumann entropy change abruptly at level
crossings as the parameters of the Breit-Rabi Hamiltonian for a hydrogen atom 
vary. An entangled state encircling the avoided crossing acquires 
the marginal Berry phase of an electron (or nuclear) spin like an eigenstate 
moving around the real crossing accumulates a Berry phase.
We have shown that the nodal points of the marginal Berry phase of 
an entangled state corresponds to the avoided crossing points.

\begin{acknowledgments}
This work is in part supported by the visitor program of Max Planck Institute 
for the Physics of Complex Systems. We would like also to thank the Binational 
Israel-US Foundation (BSF) for financial support. 
\end{acknowledgments}

\newpage

\end{document}